\documentclass[conference]{IEEEtran}
\IEEEoverridecommandlockouts
\usepackage{cite}
\usepackage{amsmath,amssymb,amsfonts}
\usepackage{algorithmic}
\usepackage{graphicx}
\usepackage{textcomp}
\usepackage{xcolor}

\newcommand{\ali}[1]{\textcolor{blue}{#1}}
\newcommand{\red}[1]{\textcolor{red}{#1}}

\newcommand{\csd}{Solana }
\newcommand{\Fig}[1]{Fig.~\ref{#1}}
\usepackage{subcaption}
\usepackage{algorithmic}
\usepackage[ruled]{algorithm2e}

\def\BibTeX{{\rm B\kern-.05em{\sc i\kern-.025em b}\kern-.08em
    T\kern-.1667em\lower.7ex\hbox{E}\kern-.125emX}}
\begin{document}

\title{In-storage Processing of I/O Intensive Applications on Computational Storage Drives 


}
\author{
 \IEEEauthorblockN{Ali HeydariGorji}
\textit{University of California Irvine}\\

\and
\IEEEauthorblockN{Mahdi Torabzadehkashi}
\IEEEauthorblockA{\textit{NGD Systems Inc.}\\ }
\vspace{4mm}
\IEEEauthorblockN{Vladimir Alves }
\IEEEauthorblockA{\textit{NGD Systems Inc.}\\}
\and
\IEEEauthorblockN{Siavash Rezaei}
\textit{NGD Systems Inc.}\\
\vspace{4mm}
\IEEEauthorblockN{Pai H. Chou }
\IEEEauthorblockA{\textit{University of California Irvine}\\}

\and
\IEEEauthorblockN{Hossein Bobarshad }
\IEEEauthorblockA{\textit{NGD Systems Inc.}\\}

}

\maketitle

\begin{abstract}

Computational storage drives (CSD) are solid-state drives (SSD) empowered by general-purpose processors that can perform in-storage processing.  They have the potential to improve both performance and energy significantly for big-data analytics by bringing compute to data, thereby eliminating costly data transfer while offering better privacy. In this work, we introduce Solana, the first-ever high-capacity(12-TB) CSD in E1.S form factor, and present an actual prototype for evaluation. To demonstrate the benefits of in-storage processing on CSD, we deploy several natural language processing (NLP) applications on datacenter-grade storage servers comprised of clusters of the Solana. Experimental results show up to 3.1x speedup in processing while reducing the energy consumption and data transfer by 67\% and 68\%, respectively, compared to regular enterprise SSDs.

\end{abstract}



\section{Introduction}
In-storage processing (ISP) is the computing paradigm of moving computation to data storage as opposed to moving data to compute engines. ISP may be the ultimate form of the near-data processing in the age of data centers, which fuel the demand for virtually unlimited storage. It has been estimated that 2.5 exabytes of data is being created every day \cite{insidebigdata_2017}, from text to images, music, and videos. The deep learning revolution has given rise to running computationally intensive algorithms on these huge amounts of data. Different technologies have been developed to support this trend towards high storage and computation demands, and they shift the bottleneck as a result.

On the storage side, hard disk drives (HDDs) have been increasing in capacity exponentially with high reliability and very competitive cost. However, HDDs have plateaued in performance, and their power consumption is ultimately lower-bounded by the mechanical motion. Solid state drives (SSD), which are based on NAND-flash memory and thus have no moving parts, offer better performance at lower power consumption, 


\begin{figure}[t]
\centerline{\includegraphics[scale=0.30]{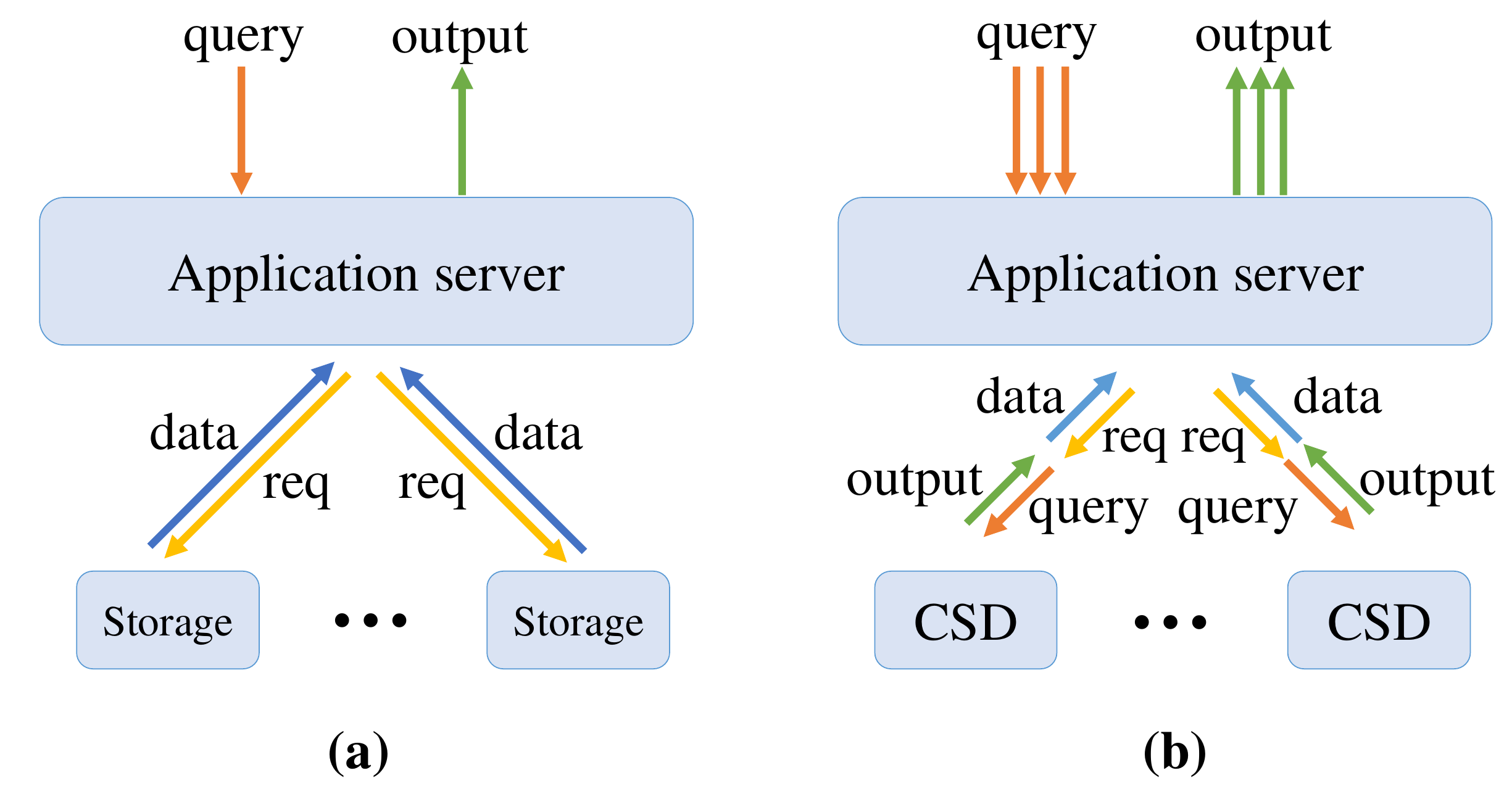}}
\caption{Handling queries on (a) a server with generic storage systems vs. (b) a server with CSDs.}
\label{sys_diag}
\end{figure}

On the computing side, general-purpose graphic processing units (GPGPU) have brought massive parallel processing to conventional PCs for running computationally intensive applications such as deep learning and crypto mining. However, GPUs are power-hungry and are not suitable for storage systems.  Data centers are now most concerned about power for not only the electric bill but also cooling, which can become another source of unreliability. As a result, GPGPUs are not so suitable for in-storage processing, but they often assume data to reside on the local computer. If the data resides on data storage, the communication network to bring it to memory will incur 5000x more energy consumption and 2500x more latency compared to loading it from volatile memory \cite{Bahn2020ImplicationsON}. 

One class of application that fits the characteristics above is natural language processing (NLP). It is a branch of artificial intelligence (AI) that enables machines to communicate with humans directly by speech or written language.
Many NLP applications use large components such as embeddings that can exceed GBs in size, or rather large input data such as audio or video input that is much much larger in size than text input \cite{dlrm}. Subsequently, the cost of data transfer between storage and host can quickly dominate in long latency and energy consumption.




In-storage processing can minimize data transfer between the host and storage by moving computation into the storage units and sending only the output back to the host. To support ISP, the storage system requires an emerging class of devices called computational storage drives (CSD), which augment the storage with processing capabilities. CSDs are being used to build new servers such as that shown in \Fig{sys_diag}, where the storage drive can serve parts of the requests.

This paper introduces a novel high-capacity CSD and demonstrates its effectiveness for ISP using some NLP applications. Our CSD prototype, code-named Solana, is a 12-TB, NAND flash SSD, equipped with a dedicated embedded processing engine along with the main SSD controller, all in \emph{E1.S form factor}. Solana's dedicated processor runs a version of embedded Linux operating system that allows users to run any Linux-based application such as Docker or distributed processes. With this CSD as the compute/storage building blocks for ISP, we show that it can effectively eliminate transfer of data out of storage unit when running our developed scheduler to distribute NLP applications onto the cluster of CSDs. We evaluate our proposed hardware/software solution for common NLP applications and also compare our design with previous state-of-the-art works on storage solutions. 

\section{Background and related work}

Near-data processing (or in this case In-Storage Processing) is a new way to break the artificial lower bound on data transmission between storage and computation by re-partitioning so that data transmission can be eliminated partially or entirely. Researchers have proposed ways of augmenting the storage with more computing power for near-data processing.\cite{scanjoin , nascent, compstor, catalina, torabzadeh_soc , biscuit , smartssd, smartssd2 , bluedbm , chapman2019computational}
Traditionally, there are two approaches to implement CSD. 

The first approach uses the same processing engine that runs the SSD controller tasks i.e. low-capability ARM M-series or R-series that yields a low performance for in-storage processing. Such design can address light computational tasks \cite{scanjoin}, but cannot be used for heavier computation tasks. 

The second approach uses an external engine coupled with the main SSD controller, to run the task computations. This method is easier to design as the storage module is separated from the ISP module. But it comes at the price of another bottleneck, that is the communication bandwidth between the storage module and the computation module. The external ISP engine can be a mixture of low-power processors such as ARM A-series or reconfigurable logic such as FPGA. FPGA-based ISP provides flexibility for implementing custom-designed accelerators. But deploying an FPGA-based unit requires RTL-level design which significantly complicates the deployment of new tasks\cite{bluedbm,smartssd2}.

Researchers in \cite{chapman2019computational} investigate a computational storage platform for big data analytics. The computation part is based on the Xilinx's Zynq, an SoC containing both ARM cores and FPGA arrays. It can directly communicate with the storage part by transferring data in P2P mode. With their software stack, it can run user binaries without modification, making it easy to port applications. Experimental results show that this system can achieve 6x improvement in query times and an average of 60\% lower CPU usage. However, no power and energy results were reported, even though the FPGAs and the off-chip accelerator are likely to be power hungry. Also, the communication between the SSD engine and the computation engine can emerge as a new bottleneck and the entire design falls behind of an integrated solution such as Solana.

A close work to ours, RecSSD \cite{recssd} proposes a near-data processing solution that improves the performance of the underlying SSD storage for embedding table operations for recommendation applications. It utilizes the internal SSD bandwidth and reduces data communication overhead between the host processor and the SSDs by offloading the entire embedding table operation, including gather and aggregation computations, to the SSDs. The hardware is a commercial OpenSSD evaluation platform\cite{openssd}. This research's main contribution is a software solution that is implemented on the flash translation layer (FTL). Experimental results show that their proposed system enhances the overall latency by up to 2x compared to off-the-shelf SSDs. Even though the results of RecSSD looks promising, the design is application specific and cannot be adapted to different applications. It also requires changes deep down to the flash translation layer (FTL) which is not an viable option for end users.

We take in-storage processing a step further by designing an ASIC that includes both the SSD controller and a dedicated ARM processor subsystem for running general applications on a single chip. The 64 bit quad-core ARM processor supports running a Linux operating system which in turn enables running executables without source-code modification or recompiling. We have also implemented a TCP/IP-based tunneling system that allows the in-storage processing engine to connect to a network, including the Internet and other processing nodes. The next section details the hardware and software that enable our system to function both as a regular storage system and as a standalone processing node.

\section{Computational Storage Platform}
\label{sec:hardware}
This section provides a high level description of the hardware prototype and software stack of the proposed Solana CSD. We explain our system with a highlight of key features for supporting efficient in-storage processing.

\subsection{CSD Hardware}
Solana is a NAND-based 12-TB storage system that uses an NVMe protocol over a 4-lane PCIe to connect to the host. The proposed architecture consists of two subsystems: flash controller unit (FCU) and in-storage processing(ISP) engine, both sharing a 6-GB Direct Random Access Memory (DRAM) through a high-speed intra-chip data bus. This section describes the architectures of FCU and ISP as well as the communication link between them.
\subsubsection{FCU Subsystem}

The proposed CSD solution should be able to act as a standard SSD with an NVMe interface, so the FCU subsystem has many parts similar to those in conventional SSDs. This subsystem includes two main modules: front-end (FE) and back-end(BE). 

The FE is responsible for receiving the IO commands from the host, checking their integrity and correctness, and interpreting them. Then, it transfers the commands to BE for execution. One of the main modules of the FE subsystem is the NVMe/PCIe interface. This interface is an ASIC module that is designed for high-performance communication with the host using NVMe over PCIe protocol.

\begin{figure}[t]
\centerline{\includegraphics[scale=0.188]{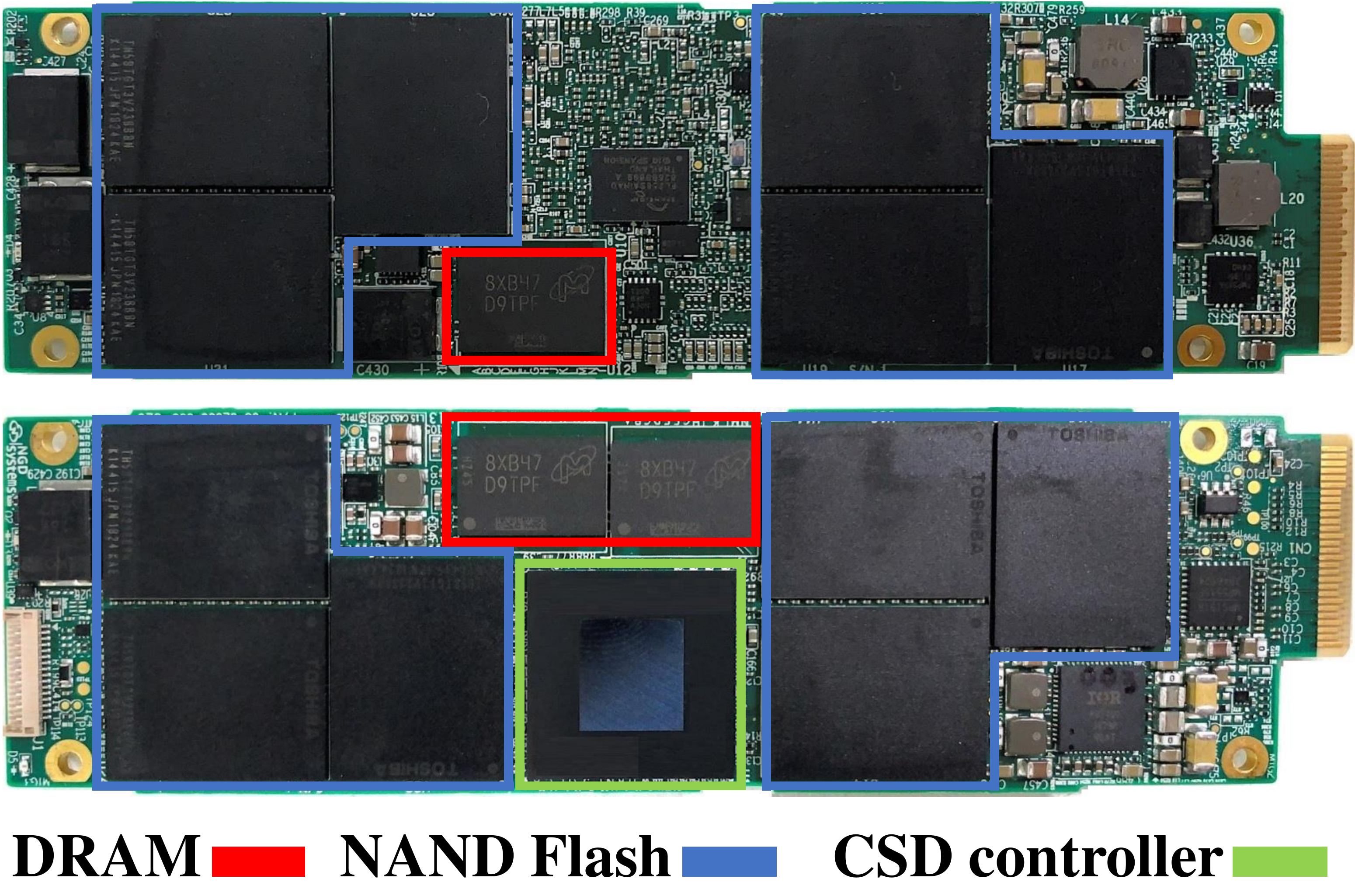}}
\caption{Solana: CSD prototype in E1.S form factor.}
\label{solana}
\end{figure}

The BE is considerably more complex than FE. It needs to communicate with the flash memory chips to complete the IO commands, and it is equipped with an error-correction unit (ECC) to remove data errors that occur in the flash memory cells. The communication between BE and the flash memory chips is handled by a 16-channel data bus. The bus is capable of doing multiple IO operations concurrently. BE is also responsible for implementing flash management routines, such as wear-leveling, address translation, and garbage collection.

Both ISP and FE are directly connected to BE. When an IO command comes from the host, FE submits it to BE. Similarly, the ISP engine can submit IO commands to BE. In other words, the ISP subsystem bypasses the FE module and the NVMe over PCIe link altogether. This provides ISP with an efficient, high-performance link to the data in the flash storage.

\subsubsection{In-Storage Processing Subsystem}

The in-storage processing  (ISP) subsystem consists of a quad-core ARM Cortex A-53 processor together with the NEON single instruction multiple data (SIMD) engines. ISP is implemented inside the same chip as the controller, and an efficient high-speed intra-chip data link connects it to the BE. Although ISP is equipped with only a quad-core processing engine, this intra-chip data link makes the ISP engine a unique processing unit. The processor also provides an environment for running different types of applications. More details of the hardware architecture are available on our previous works\cite{tos,hypertune}.

\subsection{System Software on the CSD}


We ported a full-fledged version of embedded Linux OS to our CSD to run as our ISP engine. Running a conventional Linux-based compute system makes it compatible with a wide spectrum of applications, programming languages, and command lines without modification. Linux is also extensible with our own custom features, including device drivers and file systems.

We developed a Customized Block Device Driver (CBDD) to enable accessing storage units optimized to the specific on-chip communication links/protocols, which are different from common protocols between processing units and storage devices (e.g., PCIe and SATA).  CBDD uses a command-based mechanism to communicate with the SSD controller to initiate data transfers. Using the scatter-gather mechanism, the SSD controller directly handles data transfers through the DDR addresses exposed by Linux.

\begin{figure}[t]
\centerline{\includegraphics[width=\columnwidth]{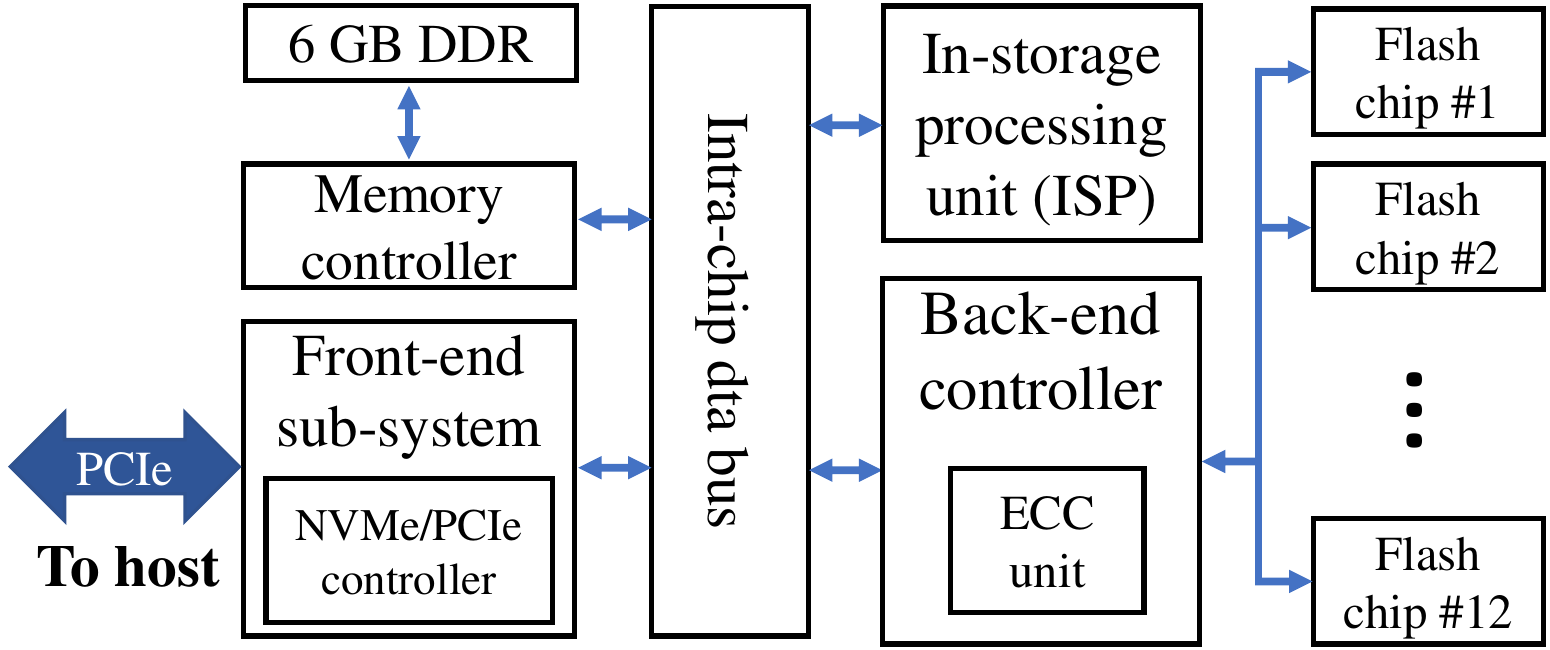}}
\caption {Hardware architecture of Solana CSD.}
\label{architecture}
\end{figure}

\begin{figure*}[t]
\centerline{\includegraphics[scale=0.7]{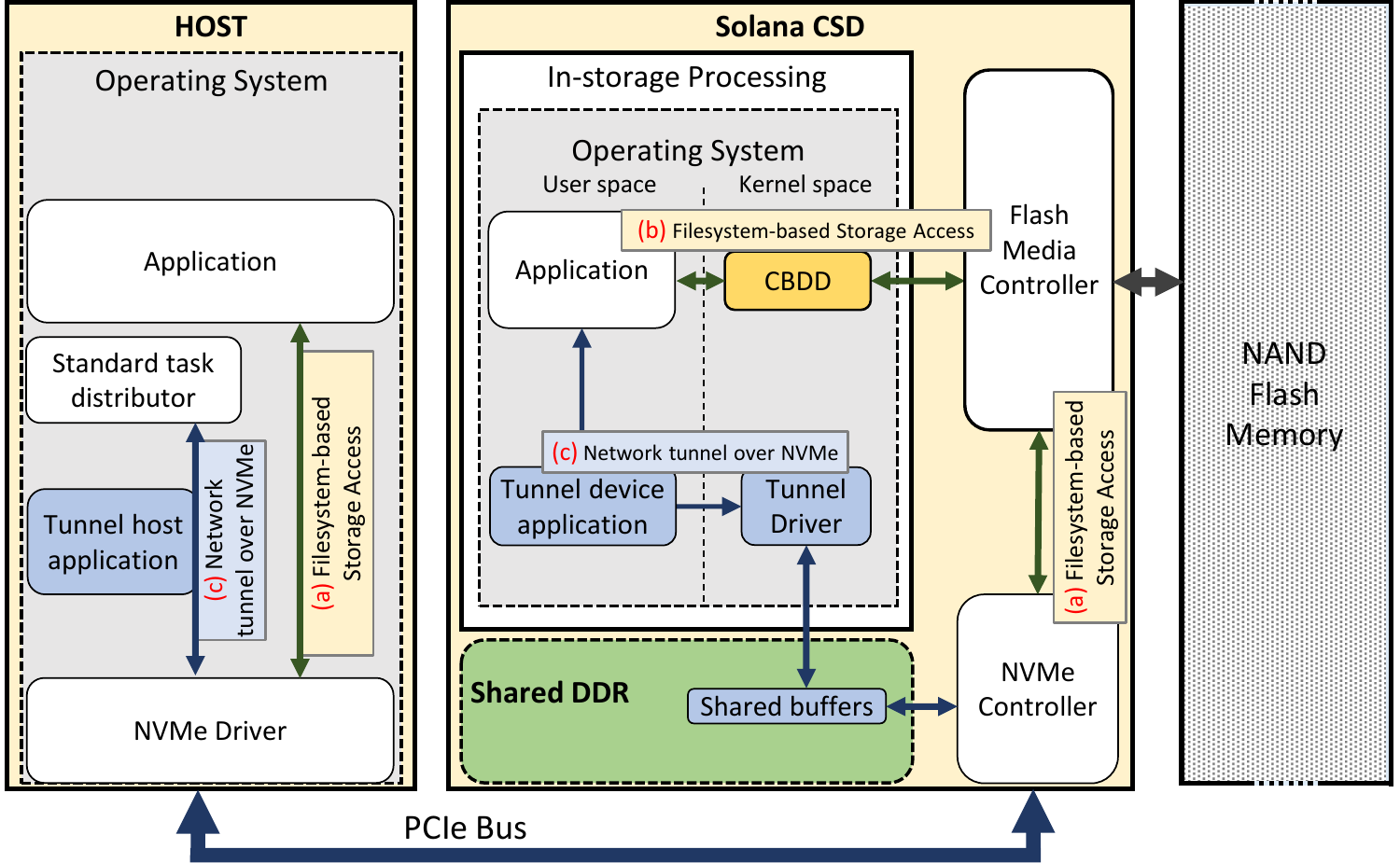}}
\caption {Solana's software stack provides three different communication paths: (a) The conventional data path through the NVMe driver to the host, (b) The path through on-chip connection to the ISP subsystem with file-system abstraction, and (c) The TCP/IP tunneling over the PCIe/NVMe.}
\label{software_stack}
\end{figure*}

One unique feature with our CSD is that the CBDD supports file-system access by the ISP applications and the host.  By allowing the embedded Linux to mount partitions on the storage and present a file API, this makes it not only easier to port applications but also more efficient to access the data\cite{hpc, torabzadeh_bigdata}.   The host CPU and ISP units can also communicate via the same partition, which is mounted by both the host and the ISP.
While conventional file systems do not allow multiple mount points, there are efficient shared file systems designed to support multiple mount points.
We use the Oracle Cluster File System (OCFS2) to enable sharing file systems and mounting the same partitions from both the host and the ISP. OCFS2 requires a TCP/IP communication link to orchestrate and update file systems of the two mounting points. Our TCP/IP tunnel provides such possibility.

\subsection{Multi-Path Software Stack}


\Fig{software_stack} depicts the different layers of the Solana's software stack. It supports three communication paths over the different interfaces on our CSD: flash-to-host, flash-to-ISP, and TCP/IP tunneling.

\subsubsection{Flash-to-Host Interface}
The conventional data path (shown as path ``a'' in \Fig{software_stack}) is through NVMe protocol over PCIe to allow the host to access the data stored in the NAND flash memory. To do so, our NVMe controller is capable of handling NVMe commands and responds to them appropriately and manages data movement through communicating with our flash media controller. 

\subsubsection{Flash-to-ISP Interface}
Path ``b'' in \Fig{software_stack} shows the data path through on-chip connection to provide file-system-based data access to the ISP engine. CBDD implements the file-system-based data access interface through communication with our flash media controller. In fact, the flash media controller is responsible for handling requests from both the ISP engine and the host. 

\subsubsection{TCP/IP tunnelling over the PCIe/NVMe}
Path ``c'' in \Fig{software_stack} shows TCP/IP communication mechanism tunneled through PCIe/NVMe to enable communication between host (and the world-wide network) and the ISP system. Such tunneling feature eliminates the need for unwieldy network setup, including cables and switches, to connect the many tightly assembled storage units, which would be impractical to maintain and scale. The proposed TCP/IP tunnel uses two shared buffers on the on-board DDR to provide the data communication. Two user-level applications, one on the host and one on the ISP running in the background, are responsible for managing NVMe-based data movement and encapsulating/decapsulating the TCP/IP data through NVMe packets.

\section{Methods and Experimental results}
\label{sec:exp_res}

We evaluate the proposed ISP system by running three NLP application on a commercial storage server. We choose the AIC FB128-LX equipped with an 8-core, 16-thread Intel\textsuperscript{\textregistered} Xeon\textsuperscript{\textregistered} Silver 4108 CPU running at 2.1~GHz and 64~GB of DRAM. It can support up to 36 E1.S drives in the front bay, 12~TB each, for a total capacity of 432~TB on a 1U class server. This section first describes our method of distributing processing loads onto the CSDs using our scheduler, followed by the results of running each benchmark.

\subsection{Methods}

To efficiently divide the processing over the host system and the CSDs with minimum overhead, we developed a scheduler that distributes applications onto multiple nodes. This scheduler is MPI-based and developed in Python. It can redirect requests or queries based on the availability of the nodes. Each node sends an ack signal to the scheduler when the current processing batch is done, which acts as a request for the next one. The scheduler runs on a separate thread on the host and wakes up every 0.2 seconds to check if there is a new message from the nodes. By putting the scheduler to sleep, the thread releases the processor and thus increases the available processing power on the host. Our setup uses the OCFS2 as the shared disk file system, which enables both the host and the ISP to access the same shared data stored on the flash\cite{stannis}. As a result, the scheduler sends only the data indexes or addresses to the ISP engine. This method significantly reduces the communication overhead by eliminating one of the most important bottlenecks in a parallel systems. It also increases the read/write speed, as all nodes access the data at a much higher speed (GBps of PCIe/NVMe for the host and DMA/hardware for the in-situ vs.\ MBps of TCP/IP) by communicating with the flash directly.

\begin{figure*}[t]
\centerline{\includegraphics[width=\textwidth]{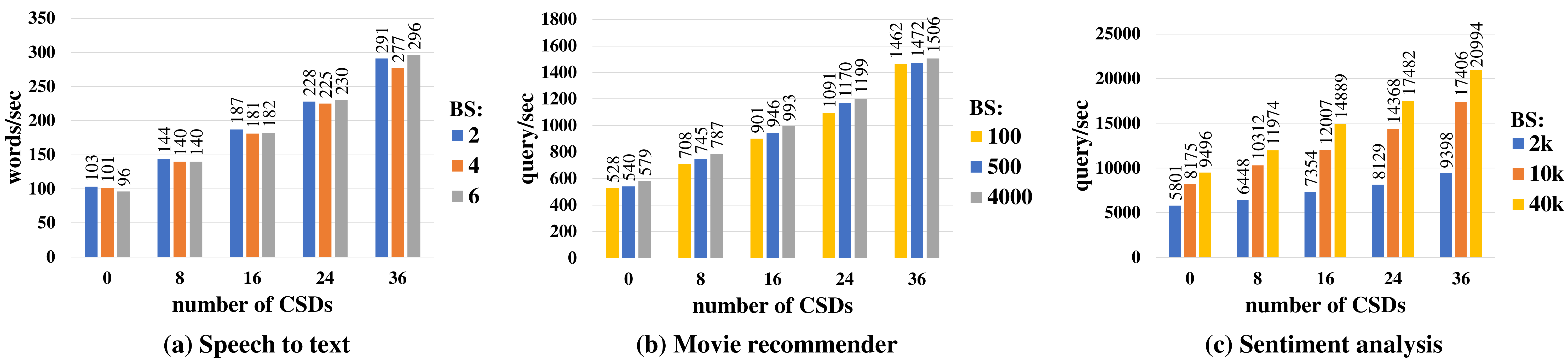}}
\caption {Experimental results for three NLP benchmarks and different batchsizes.}
\label{result}
\end{figure*}

Two important parameters for the scheduler are the \emph{batch size}, or the size of the chunk of data that is assigned to each node at a time, and the \emph{batch ratio}, which is the ratio between the batch size for the host and the ones for the CSDs. The \emph{batch ratio} is determined based on the difference in the processing power in a heterogeneous system. Since the host has a more powerful CPU (Xeon) than the CSD does (ARM A53), the \emph{batch ratio} is considerably large, ranging from 20 to 30 based on the experimental results. Any ratio other than the optimal \emph{batch ratio} results in under-utilization of the system. Using the \emph{batch ratio} decreases the workload of the scheduler and also increases the host's performance due to lower scheduling overhead and larger chunks of data to process at a time. The batch size for each application can be determined by running a small test to obtain the best range for the batch size. In the following subsection, we will investigate the performance of different configurations on our prototype. Note that due to unavailability, we cannot compare  our prototype with other CSD projects and hence, we can just compare it with a baseline system that uses the CSD only as an storage drive.

\subsection{Results}

\subsubsection{Speech to text benchmark}
This benchmark is developed based on Vosk, an offline speech recognition toolkit \cite{vosk}. Our dataset is LJ Dataset speech \cite{ljspeech17}, a public-domain speech dataset consisting of 13,100 short audio clips. The dataset is about 24 hours long and consists of 225,715 words. By running a small benchmark on single nodes, we get 102 words/sec and 5.3 words/sec processing speed for the host and the CSDs, respectively. That yields an approximate batch size ratio of 20, which we will use in the scheduler. This ratio is almost the same for all three applications.

\Fig{result}(a) shows the overall output of this benchmark in terms of the number of transcribed words per second, based on the batch size and the number of engaged CSDs. By augmenting the host with the processing power of all 36 CSDs, the output rate increases from 96 words/sec to 296 words/sec for batch size of 6. That is 3.1x better performance compared to the host alone. \Fig{result}(a) also shows that the processing speed does not change much (less than 7\%) when varying the batch size. In terms of IO transfers, CSD engines processed 68\% of the input data ($(296-96) \div 296\simeq0.68$), which means that 2.58~GB out of the 3.8~GB of the dataset never left the storage unit, and the only output of about 1.2~MB that was transferred to the host was the output text. This reduction in number of IO's and in data transfer size drastically reduces the power consumption, network congestion, and use of the host's CPU memory bandwidth. It also enhances the security and the privacy aspects of the data as it never leaves the storage device.

\subsubsection{Movie Recommender Benchmark}
Our second benchmark is a movie recommendation system based on \cite{recomm}. The recommender creates a new metadata entry for each movie, including the title, genres, director, main actors, and the story line keywords. It then uses the cosine similarity function to generate a similarity matrix that will be used later to suggest contents. As an extra step, the ratings and the popularity indexes will be used to filter the top results. For each query, the target movie title is sent to the recommender function that returns the top-10 similar movies. We used the MovieLens Dataset \cite{movie}, which consists of 27 million ratings and 1.1 million tag applications applied to 58,000 movie titles by 280,000 users. We ran the training process once and stored the matrix on flash for further uses. To simulate the queries, we made a list of all movie titles and randomly shuffled them into a new list. \Fig{result}(b) shows the results for movie recommendation queries with different configurations. With 36 drives, the system can address 1506 queries/second compared to the 579 queries/second of the standalone host, which is 2.6x improvement. Just like the speech-to-text benchmark, changing the batch size does not significantly affect the outcome by more than 3\%.

\subsubsection{Sentiment Analysis Benchmark}

\begin{figure}[t]
\centerline{\includegraphics[scale=0.41]{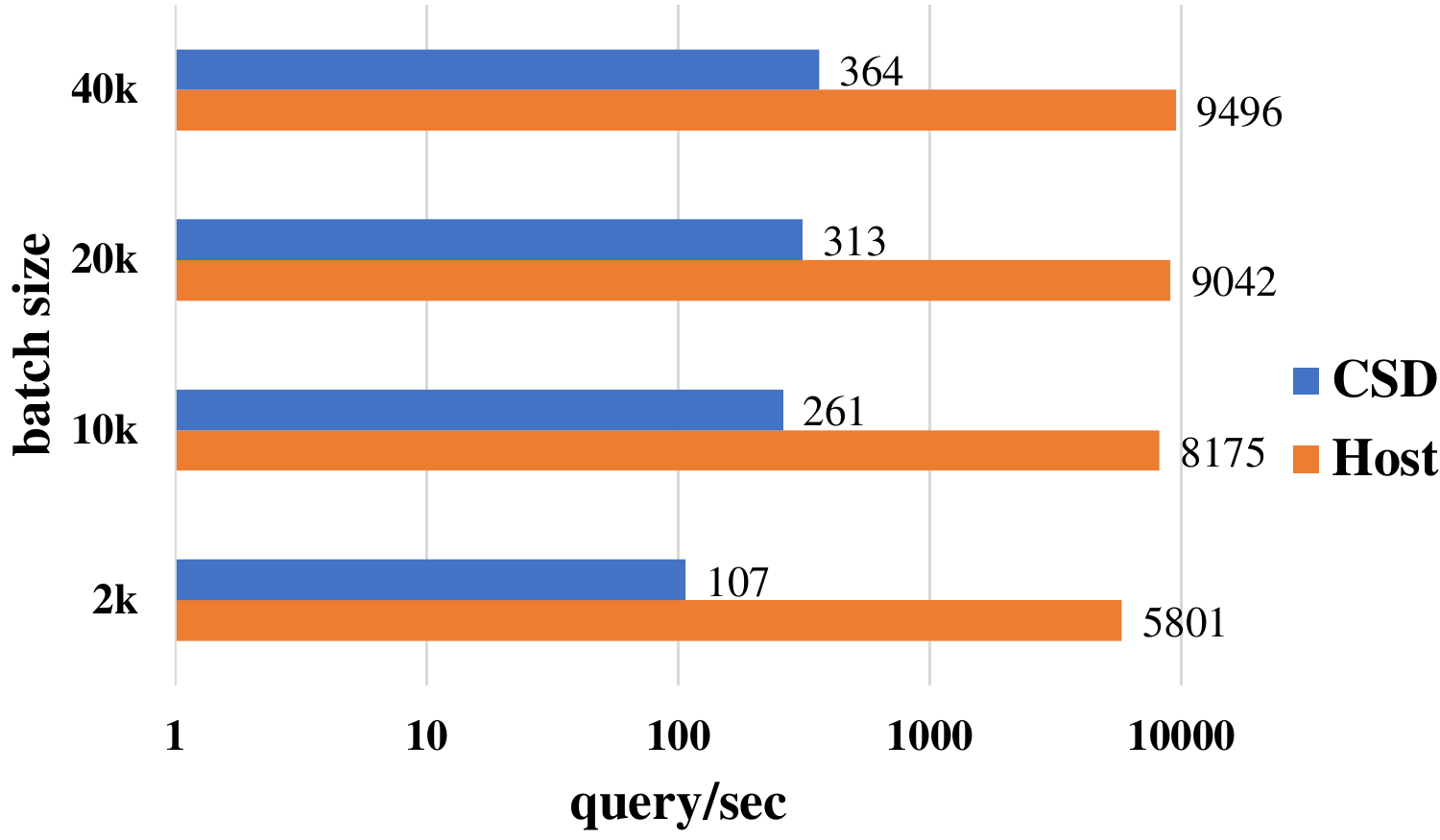}}
\caption{1-node performance for sentiment analysis benchmark.}
\label{singlenode_sentiment}
\end{figure}

The last benchmark is a Twitter sentiment analysis app based on Python's Natural Language Toolkit (NLTK), depicted from \cite{sentiment_anal} and modified to fit to our parallelization goals. It uses labeled data to train a model to detect the positivity or negativity of the tweet's content and then uses the model to predict the sentiment of the incoming tweets. Our test dataset consists of 1.6 million tweets \cite{1.6tweet} and are duplicated in cases where we need a larger number of queries. We ran a single node test to evaluate the host and the CSDs performance for different batch sizes. Unlike the other two benchmarks, performance changes considerably based on the batch size. \Fig{singlenode_sentiment} shows the performance in terms of queries per second for different batch sizes on a logarithmic scale. It shows that as batch sizes increase, both {\csd} and the host get better performance but also increased latency in processing some of the queries.  Because the input are sequentially fed to the nodes, once a batch of queries is assigned to one processing node, each query has to wait for prior queries in the same batch to finish but cannot be migrated to some other idle nodes; in smaller batch sizes, different batches of queries could be assigned to other available nodes and therefore the total wait time is less. Based on these numbers, we set the batch size ratio to $9496\div364\simeq26$ and ran the benchmark with 8 million tweets. \Fig{result}(c) shows the performance for different batch sizes and on different numbers of CSDs. The best result is for the batch size of 40k, where the number of queries per second increases from 9496 to 20994, or 2.2x the performance.

\subsection{Power and Energy Analysis}

\begin{figure}[t]
\centerline{\includegraphics[scale=0.42]{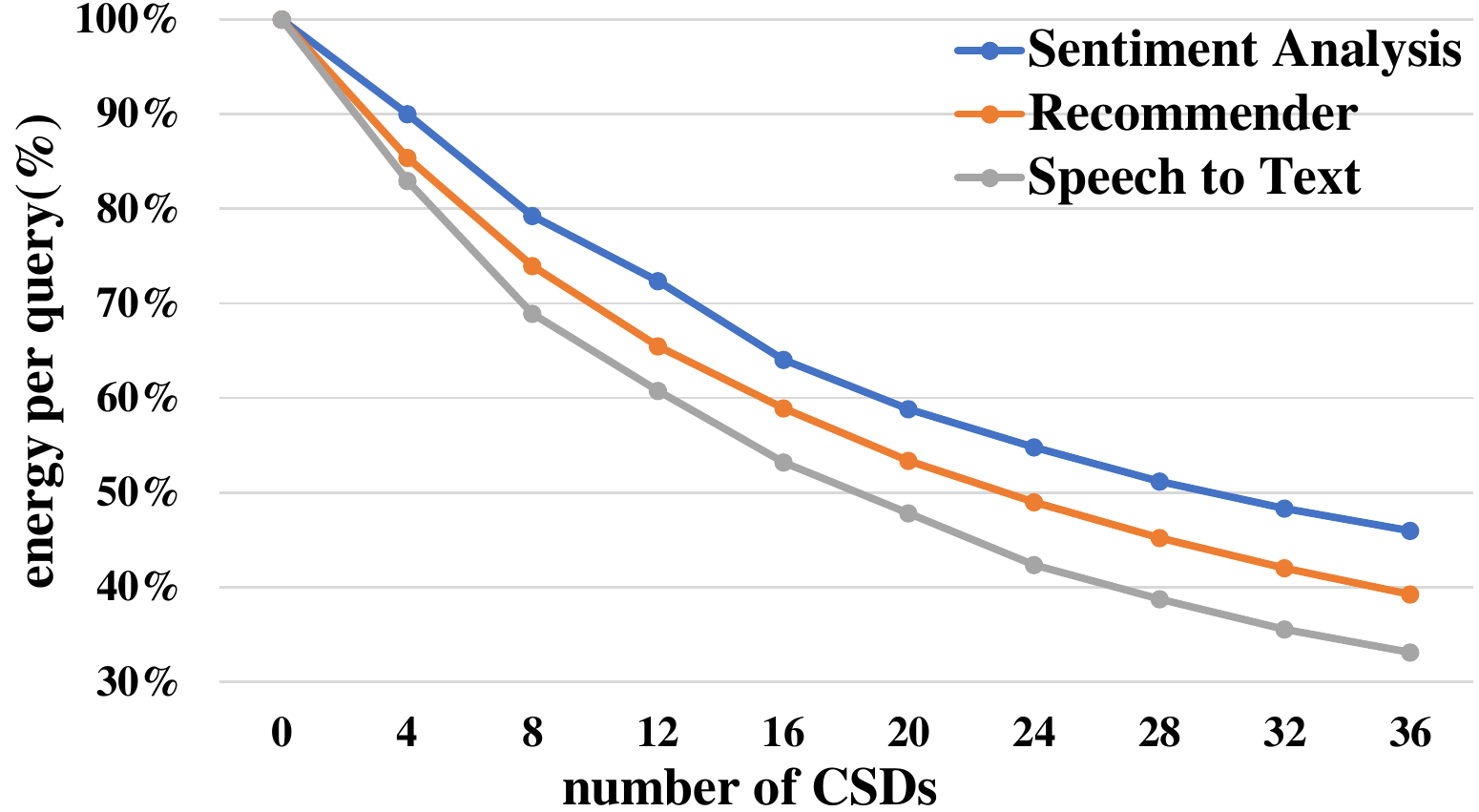}}
\caption{Energy per query, normalized to host-only setup.}
\label{energy_percent}
\end{figure}

We measured the power and energy consumption of the server using an HPM-100A power meter. The overall power consumption includes the host processor, the storage systems, and peripherals including the cooling system. Since there is no comparable E1.S drive on the market with 12~TB capacity, we choose the baseline test system to be the same server but with the ISP engines disabled. In idle mode, the server consumes 167~W without storage drives, or 405~W with 36 CSDs. Thus, each CSD consumes an average of 6.6~W, compared to 10-15~W for commercial E1.S units with lower storage capacity in normal operation mode. When we ran the benchmarks, the power consumption of the entire system was up to 482~W without enabling ISP (i.e., CSD acting as storage only), compared to 492~W with all 36 ISP engines running. That is, each ISP engine consumes 0.28~W on top of the storage-only feature. \Fig{energy_percent} shows the normalized energy consumption per query, or per word in the case of the text-to-speech benchmark. We used a normalized diagram to better show the energy saving trend. Table~\ref{tab:table_1} summarizes the measured energy results along with a summary of the experiments.

\begin{table}[t]
\caption{Summary of the experimental results.}
\label{tab:table_1}
\resizebox{\columnwidth}{!}
{
\begin{tabular}{lccc}
\hline
\textbf{application}                    & \parbox{4em}{\centering speech to text} & \parbox{4em}{\centering recom- mender} & \parbox{4em}{\centering sentiment analysis}          \\ \hline\hline
\textbf{dataset size}                     & 226K             & 58K             & 1.6M           \\ \hline
\textbf{output accuracy}                         & same             & same            & same           \\ \hline
\textbf{max speedup}                      & 3.1x             & 2.8x            & 2.2x           \\ \hline
\textbf{energy per query (host) (mJ)}     & 5021             & 832             & 51             \\ \hline
\textbf{energy per query (w/CSD) (mJ)}    & 1662             & 327             & 23             \\ \hline
\textbf{energy saving per query (\%)}               & 67\%             & 61\%            & 54\%           \\ \hline
\textbf{data processed on host (\%)}  & 32\%             & 36\%            & 44\%           \\ \hline
\textbf{data processed in CSDs (\%)}  & 68\%             & 64\%            & 56\%           \\ \hline
\end{tabular}
}
\end{table}

\section{Conclusions}

SSDs not only brought fast, power-efficient data storage but also new opportunities for a new class of devices in the form of computational storage drives (CSD) that can support near-data (or in-storage) processing. Moving computation to storage not only eliminates most of the data transfers needed and saves the time and energy, but also has the added benefit of better privacy. All these benefits come at negligible cost, as the cost of the storage system is dominated by the NAND chip. 

In this work, we introduced Solana, the first  12TB Computational Storage Drive in \emph{E1.S} form factor. Unlike other CSDs, Solana has an integrated In-Storage Processing engine with the SSD controller on one chip, makes it a true ISP capable drive. We demonstrated the effectiveness of deploying our proposed design on storage servers which can achieve up to 3.1x speedup for a number of common NLP applications, while reducing the energy consumption and data transfer by 67\% and 68\%, respectively, compared to regular enterprise SSDs.  For future work, we plan to develop a data-aware distributed system that can benefit not only from temporal locality but also from spacial locality of data, by  classifying queries into categorical groups and redirecting them to associated nodes.

\bibliographystyle{IEEEtran}
\bibliography{5-bibliography}

\begin{thebibliography}{10}
\providecommand{\url}[1]{#1}
\csname url@samestyle\endcsname
\providecommand{\newblock}{\relax}
\providecommand{\bibinfo}[2]{#2}
\providecommand{\BIBentrySTDinterwordspacing}{\spaceskip=0pt\relax}
\providecommand{\BIBentryALTinterwordstretchfactor}{4}
\providecommand{\BIBentryALTinterwordspacing}{\spaceskip=\fontdimen2\font plus
\BIBentryALTinterwordstretchfactor\fontdimen3\font minus
  \fontdimen4\font\relax}
\providecommand{\BIBforeignlanguage}[2]{{%
\expandafter\ifx\csname l@#1\endcsname\relax
\typeout{** WARNING: IEEEtran.bst: No hyphenation pattern has been}%
\typeout{** loaded for the language `#1'. Using the pattern for}%
\typeout{** the default language instead.}%
\else
\language=\csname l@#1\endcsname
\fi
#2}}
\providecommand{\BIBdecl}{\relax}
\BIBdecl

\bibitem{insidebigdata_2017}
\BIBentryALTinterwordspacing
``How data is stored and what we do with it,'' Nov 2017. [Online]. Available:
  \url{https://insidebigdata.com/2017/11/12/how-data-is-stored-and-what-we-do-with-it/}
\BIBentrySTDinterwordspacing

\bibitem{Bahn2020ImplicationsON}
H.~Bahn and K.~Cho, ``Implications of {NVM} based storage on memory subsystem
  management,'' \emph{Applied Sciences}, vol.~10, p. 999, 2020.

\bibitem{dlrm}
M.~Naumov, D.~Mudigere, H.-J.~M. Shi, J.~Huang, N.~Sundaraman, J.~Park,
  X.~Wang, U.~Gupta, C.-J. Wu, A.~G. Azzolini, D.~Dzhulgakov, A.~Mallevich,
  I.~Cherniavskii, Y.~Lu, R.~Krishnamoorthi, A.~Yu, V.~Kondratenko, S.~Pereira,
  X.~Chen, W.~Chen, V.~Rao, B.~Jia, L.~Xiong, and M.~Smelyanskiy, ``Deep
  learning recommendation model for personalization and recommendation
  systems,'' 2019.

\bibitem{scanjoin}
\BIBentryALTinterwordspacing
S.~Kim, H.~Oh, C.~Park, S.~Cho, S.-W. Lee, and B.~Moon, ``In-storage processing
  of database scans and joins,'' \emph{Inf. Sci.}, vol. 327, no.~C, p.
  183–200, Jan. 2016. [Online]. Available:
  \url{https://doi.org/10.1016/j.ins.2015.07.056}
\BIBentrySTDinterwordspacing

\bibitem{nascent}
\BIBentryALTinterwordspacing
S.~Salamat, A.~Haj~Aboutalebi, B.~Khaleghi, J.~H. Lee, Y.~S. Ki, and T.~Rosing,
  ``Nascent: Near-storage acceleration of database sort on smartssd,'' in
  \emph{The 2021 ACM/SIGDA International Symposium on Field-Programmable Gate
  Arrays}, ser. FPGA '21.\hskip 1em plus 0.5em minus 0.4em\relax New York, NY,
  USA: Association for Computing Machinery, 2021, p. 262–272. [Online].
  Available: \url{https://doi.org/10.1145/3431920.3439298}
\BIBentrySTDinterwordspacing

\bibitem{compstor}
M.~Torabzadehkashi, S.~Rezaei, V.~Alves, and N.~Bagherzadeh, ``Compstor: An
  in-storage computation platform for scalable distributed processing,'' in
  \emph{2018 IEEE International Parallel and Distributed Processing Symposium
  Workshops (IPDPSW)}.\hskip 1em plus 0.5em minus 0.4em\relax IEEE, 2018, pp.
  1260--1267.

\bibitem{catalina}
M.~Torabzadehkashi, S.~Rezaei, A.~Heydarigorji, H.~Bobarshad, V.~Alves, and
  N.~Bagherzadeh, ``Catalina: In-storage processing acceleration for scalable
  big data analytics,'' in \emph{2019 27th Euromicro International Conference
  on Parallel, Distributed and Network-Based Processing (PDP)}, 2019, pp.
  430--437.

\bibitem{torabzadeh_soc}
M.~Torabzadehkashi, \emph{SoC-Based In-Storage Processing: Bringing Flexibility
  and Efficiency to Near-Data Processing}.\hskip 1em plus 0.5em minus
  0.4em\relax University of California, Irvine, 2019.

\bibitem{biscuit}
B.~Gu, A.~S. Yoon, D.-H. Bae, I.~Jo, J.~Lee, J.~Yoon, J.-U. Kang, M.~Kwon,
  C.~Yoon, S.~Cho, J.~Jeong, and D.~Chang, ``Biscuit: A framework for near-data
  processing of big data workloads,'' in \emph{2016 ACM/IEEE 43rd Annual
  International Symposium on Computer Architecture (ISCA)}, 2016, pp. 153--165.

\bibitem{smartssd}
J.~H. Lee, H.~Zhang, V.~Lagrange, P.~Krishnamoorthy, X.~Zhao, and Y.~S. Ki,
  ``Smartssd: Fpga accelerated near-storage data analytics on ssd,'' \emph{IEEE
  Computer Architecture Letters}, vol.~19, no.~2, pp. 110--113, 2020.

\bibitem{smartssd2}
M.~Soltaniyeh, V.~L. Moutinho Dos~Reis, M.~Bryson, R.~Martin, and
  S.~Nagarakatte, ``Near-storage acceleration of database query processing with
  smartssds,'' in \emph{2021 IEEE 29th Annual International Symposium on
  Field-Programmable Custom Computing Machines (FCCM)}, 2021, pp. 265--265.

\bibitem{bluedbm}
\BIBentryALTinterwordspacing
S.-W. Jun, M.~Liu, S.~Lee, J.~Hicks, J.~Ankcorn, M.~King, S.~Xu, and Arvind,
  ``{BlueDBM}: An appliance for big data analytics,'' \emph{SIGARCH Comput.
  Archit. News}, vol.~43, no.~3S, p. 1–13, Jun. 2015. [Online]. Available:
  \url{https://doi.org/10.1145/2872887.2750412}
\BIBentrySTDinterwordspacing

\bibitem{chapman2019computational}
K.~Chapman, M.~Nik, B.~Robatmili, S.~Mirkhani, and M.~Lavasani, ``Computational
  storage for big data analytics,'' in \emph{Proceedings of 10th International
  Workshop on Accelerating Analytics and Data Management Systems (ADMS’19)},
  2019.

\bibitem{recssd}
\BIBentryALTinterwordspacing
M.~Wilkening, U.~Gupta, S.~Hsia, C.~Trippel, C.-J. Wu, D.~Brooks, and G.-Y.
  Wei, \emph{{RecSSD}: Near Data Processing for Solid State Drive Based
  Recommendation Inference}.\hskip 1em plus 0.5em minus 0.4em\relax New York,
  NY, USA: Association for Computing Machinery, 2021, p. 717–729. [Online].
  Available: \url{https://doi.org/10.1145/3445814.3446763}
\BIBentrySTDinterwordspacing

\bibitem{openssd}
\BIBentryALTinterwordspacing
 [Online]. Available: \url{http://www.openssd.io/}
\BIBentrySTDinterwordspacing

\bibitem{tos}
\BIBentryALTinterwordspacing
J.~Do, V.~C. Ferreira, H.~Bobarshad, M.~Torabzadehkashi, S.~Rezaei,
  A.~Heydarigorji, D.~Souza, B.~F. Goldstein, L.~Santiago, M.~S. Kim, P.~M.~V.
  Lima, F.~M.~G. Fran\c{c}a, and V.~Alves, ``Cost-effective, energy-efficient,
  and scalable storage computing for large-scale ai applications,'' \emph{ACM
  Trans. Storage}, vol.~16, no.~4, oct 2020. [Online]. Available:
  \url{https://doi.org/10.1145/3415580}
\BIBentrySTDinterwordspacing

\bibitem{hypertune}
A.~HeydariGorji, S.~Rezaei, M.~Torabzadehkashi, H.~Bobarshad, V.~Alves, and
  P.~H. Chou, ``Hypertune: Dynamic hyperparameter tuning for efficient
  distribution of dnn training over heterogeneous systems,'' in \emph{2020
  IEEE/ACM International Conference On Computer Aided Design (ICCAD)}, 2020,
  pp. 1--8.

\bibitem{hpc}
M.~Torabzadehkashi, A.~Heydarigorji, S.~Rezaei, H.~Bobarshad, V.~Alves, and
  N.~Bagherzadeh, ``Accelerating hpc applications using computational storage
  devices,'' in \emph{2019 IEEE 21st International Conference on High
  Performance Computing and Communications; IEEE 17th International Conference
  on Smart City; IEEE 5th International Conference on Data Science and Systems
  (HPCC/SmartCity/DSS)}.\hskip 1em plus 0.5em minus 0.4em\relax IEEE, 2019, pp.
  1878--1885.

\bibitem{torabzadeh_bigdata}
M.~Torabzadehkashi, S.~Rezaei, A.~HeydariGorji, H.~Bobarshad, V.~Alves, and
  N.~Bagherzadeh, ``Computational storage: an efficient and scalable platform
  for big data and hpc applications,'' \emph{Journal of Big Data}, vol.~6,
  no.~1, pp. 1--29, 2019.

\bibitem{stannis}
A.~HeydariGorji, M.~Torabzadehkashi, S.~Rezaei, H.~Bobarshad, V.~Alves, and
  P.~H. Chou, ``Stannis: Low-power acceleration of dnn training using
  computational storage devices,'' in \emph{2020 57th ACM/IEEE Design
  Automation Conference (DAC)}, 2020, pp. 1--6.

\bibitem{vosk}
\BIBentryALTinterwordspacing
 [Online]. Available: \url{https://alphacephei.com/vosk/}
\BIBentrySTDinterwordspacing

\bibitem{ljspeech17}
K.~Ito and L.~Johnson, ``The {LJ} speech dataset,''
  \url{https://keithito.com/LJ-Speech-Dataset/}, 2017.

\bibitem{recomm}
\BIBentryALTinterwordspacing
J.~Ng, ``Content-based recommender using natural language processing ({NLP}),''
  Apr 2020. [Online]. Available:
  \url{https://towardsdatascience.com/content-based-recommender-using-natural-language-processing-nlp-159d0925a649}
\BIBentrySTDinterwordspacing

\bibitem{movie}
\BIBentryALTinterwordspacing
``Movielens,'' Mar 2021. [Online]. Available:
  \url{https://grouplens.org/datasets/movielens/}
\BIBentrySTDinterwordspacing

\bibitem{sentiment_anal}
\BIBentryALTinterwordspacing
DigitalOcean, ``How to perform sentiment analysis in {Python} 3 using the
  natural language toolkit ({NLTK}),'' Jan 2021. [Online]. Available:
  \url{https://www.digitalocean.com/community/tutorials/how-to-perform-sentiment-analysis-in-python-3-using-the-natural-language-toolkit-nltk}
\BIBentrySTDinterwordspacing

\bibitem{1.6tweet}
A.~Go, R.~Bhayani, and L.~Huang, ``Twitter sentiment classification using
  distant supervision,'' \emph{Processing}, vol. 150, 01 2009.

\end{thebibliography}

\end{document}